# Low-Cost Cross-Correlation Noise Setup for Measuring the Boltzmann Constant and the Elementary Charge

(Dated: February 15, 2026)

Zitong Peng[1]

College of Physics and Electronic Engineering, Sichuan Normal University, Chengdu 610101, People's Republic of China

Jie Zheng[2]

College of Physics and Electronic Engineering, Sichuan Normal University, Chengdu 610101, People's Republic of China

Xiaokai Yue*

College of Physics and Electronic Engineering, Sichuan Normal University, Chengdu 610101, People's Republic of China

Abstract

We present a low-cost experimental setup to measure the Boltzmann constant ($k_\text{B}$) and the elementary charge ($e$) through thermal and shot noise, achieving relative accuracies of up to 1%. The system utilizes a cost-effective ADC module integrated with a carefully selected low-noise audio operational amplifier, resulting in a simple and compact circuit design that better satisfies experimental requirements. Furthermore, the approach employs a dual-channel sampling method to measure cross-correlation noise, thereby significantly reducing the uncorrelated background noise of the device and improving the signal-to-noise ratio. The noise measurement methodology offers an intuitive demonstration of microscopic fluctuations, making it a valuable teaching tool. In addition, the experimental setup employs standard electronic components, is low-cost and robust, making it well-suited for widespread adoption in university and even high school laboratories.

# I. INTRODUCTION

The measurement of the Boltzmann constant ($k_B$) and the elementary charge ($e$) through electrical noise in circuits has been employed in educational settings since the mid-20$^{th}$ century [1]. Early educational implementations, such as the undergraduate noise thermometry experiment by Kittel et al. [2], demonstrated how standard electronic instruments could be used to measure temperature. This experimental approach, which utilizes microscopic fluctuations to determine fundamental physical constants, provides a robust illustration of the broad applicability of these concepts in the mathematical modeling of random processes. In conventional noise measurement experiments, oscilloscopes are typically employed [1–4], enabling students to simultaneously observe both time-domain fluctuations and the frequency-domain noise power spectrum. However, the noise signal to be measured is typically on the order of nV/$\sqrt{\text{Hz}}$ or even smaller. Conventional time-domain voltage acquisition devices often exhibit insufficient resolution for such direct measurements. Therefore, conducting this experiment necessitates the use of either high-precision acquisition equipment or a high-gain preamplifier. This significantly increases the experimental cost, hindering its widespread adoption. With the development of the electronics industry, to reduce costs and enhance accessibility for educational purposes, References [5–7] introduce an alternative approach that replaces the original design with a combination of a low-noise amplifier and an RMS-DC converter to indirectly measure the noise spectrum. The disadvantage brought about by this solution is that students cannot directly observe the noise power spectrum distribution. Furthermore, the method is highly susceptible to interference signals within the bandwidth, which can cause the inherent background noise of these instruments to mask the target signal; additionally, the circuit implementation is relatively complex, posing a significant challenge for students attempting independent work.

To address these limitations, we have developed an experimental setup that is low-cost and has the intuitive approach. The system combines Burr-Brown™ audio operational amplifiers with a low-sample-rate ADC module, offering a simple circuit architecture well-suited for educational use. The amplifier's FET-input structure provides low voltage and current noise density and a very low 1/f corner frequency. Replacing the oscilloscope with an inexpensive ADC module further simplifies the experimental procedure. This design enables students to directly observe noise spectrum characteristics, deepen their physical understanding, and identify interference signals within the measurement bandwidth, thereby supporting more effective error analysis and experimental optimization. The overall setup is low-cost, user-friendly, and readily adaptable for do-it-yourself (DIY) experimentation. Furthermore, we employ cross-correlation technique, which is a well-established method for extracting weak signals embedded in noise. Its development can be traced to early implementations such as Fink's π-network noise thermometer, which utilized voltage correlation to achieve temperature measurements [8]. Later refinements, including the two-amplifier correlation method described by Pellegrini et al., significantly improved the accuracy of noise estimates [9]. It has also been adopted in undergraduate laboratories for determining fundamental constants and has proven valuable in various educational physics contexts [10–12]. By incorporating cross-correlation measurements, our experiment not only suppresses instrumental background noise effectively but also offers students insight into the underlying principles, thereby enriching its pedagogical content. Building on these prior applications, we use cross-correlation to further suppress instrumental noise in the measurement of the fundamental constants $k_B$ and $e$, enhancing both the accuracy and educational value of the experiment. This approach also helps stimulate student interest in advanced topics such as quantum transport [13] and two-photon statistics [14].

## II. THEORY

### A. Power Spectral Density

The power spectral density (PSD) is a fundamental quantity that describes how the power of a signal is distributed as a function of frequency [15–17]. In physical and engineering contexts, the PSD provides a powerful frequency-domain characterization of random fluctuations. One of its most common applications is to analyze the noise in electronic circuits [17]. For a continuous-time voltage signal $V(t)$, the voltage noise PSD $S_V(f)$ specifies the mean-square voltage fluctuation per unit bandwidth around frequency $f$. For a signal observed over a finite duration $t_f$, $S_V(f)$ can be estimated from the squared magnitude of its Fourier transform

$$S_V(f) = \frac{2}{t_f} \left| \int_0^{t_f} e^{-i\omega t} V(t) dt \right|^2, \qquad (1)$$

where $\omega = 2\pi f$ and we use the Fourier transform definition without a prefactor. The physical meaning of $S_V(f)$ is straightforward. For a small frequency interval $\Delta f$, the power dissipated in a resistor $R$ is $\Delta P = S_V(f) \Delta f / R$. Thus, the PSD directly links voltage fluctuations to measurable power.

### B. The Fourier Transform

Applying the Fourier Transform to voltage data enables the calculation of the power spectral density. The Fourier transform $V(\omega)$ of the voltage signal $V(t)$ is defined as: $V(\omega) = \int e^{-i\omega t} V(t) dt$. We adopt the non-unitary Fourier transform convention, consistent with standard treatments of noise spectra. The data collected by the ADC are discrete. Therefore, the Discrete Fourier Transform (DFT) is employed to calculate the frequency-domain voltage. In practice, this is implemented via the efficient Fast Fourier Transform (FFT) algorithm. Given $N_t$ samples acquired over a duration $t_f$, the time domain and the angular frequency domain can be discretized

$$\begin{cases} t_n = \dfrac{t_f}{N_t} n, \; n = [0, 1, \ldots, N_t - 1] \\ \omega_k = \dfrac{2\pi}{t_f} k, \; k = \left[0, 1, \ldots, \dfrac{N_t}{2} - 1\right] \end{cases}. \qquad (2)$$

And the DFT of the voltage data is given by

$$V(f_k)=\sum_{n=0}^{N_t-1}\frac{t_f}{N_t}e^{-\frac{i2\pi nk}{N_t}}V_n(t)=\frac{t_f}{N_t}\text{DFT}(\mathbf{V})_k, \qquad (3)$$

where $f_k=k/t_f$ is the physical frequency value of the $k^{th}$ point in the discrete spectrum. When configured to a sampling rate of $f_s$, the ADC acquires $N_t$ data points per trace. This corresponds to a sampling time of $t_f=N_t/f_s$. The power spectral density is then computed for each acquired dataset. The experimental accuracy can be improved by taking repeated measurements and averaging the results. The detailed process is described in Section IV-A. The noise spectrum is then given by

$$S_V(f_k)=2\frac{\langle|V(f_k)|^2\rangle}{t_f}=2\frac{t_f}{N_t^2}\langle|\text{DFT}(\mathbf{V})_k|^2\rangle, \qquad (4)$$

where $\langle\ldots\rangle$ represents average over repeated experiments. The Fourier transform plays a fundamental role in the undergraduate physics curriculum and Reference [18] provides a comprehensive and clear guide on how to program in a computer.

**C. Cross-correlation Noise**

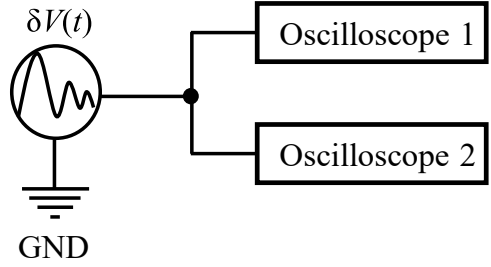

Fig. 1: Schematic diagram of cross-correlation noise measurement.

Given two signals $V_1(t)$ and $V_2(t)$, each of which possess power spectral densities $S_1(f)$ and $S_2(f)$, it is possible to define a cross power spectral density $S_{\text{cross}}(f)$ [19]. Its basic principle is as follows: if two quantities A and B are uncorrelated and each has zero mean $\langle A\rangle=\langle B\rangle=0$, then the expected value of their product is zero, $\langle AB\rangle=0$. As shown in the Fig. 1, the voltage signal measured by Oscilloscope 1 comprises two components, $V_1(t)=\delta V(t)+V_{\text{osc1}}(t)$, where $\delta V(t)$ is the noise of interest and $V_{\text{osc1}}(t)$ represents the background noise contributed by the Oscilloscope 1. Thus, the measured single-channel noise power spectrum is given by

$$\begin{aligned}S_1(f)&=\frac{2}{t_f}\langle|\delta V(f)|^2+|V_{\text{osc}}(f)|^2+\delta V(f)V_{\text{osc1}}^*(f)+V_{\text{osc1}}(f)\delta V(f)^*\rangle\\ &=S_{\delta V}(f)+S_{\text{osc1}}(f),\end{aligned} \qquad (5)$$

where $\langle\ldots\rangle$ denotes long-term averaging, $\delta V(f)$ and $V_{\text{osc1}}(f)$ are the Fourier transforms of $\delta V(t)$ and $V_{\text{osc1}}(t)$ respectively. Since $\delta V(t)$ and $V_{\text{osc1}}(t)$ are uncorrelated, the expected values of the last two terms in the first step of the derivation are zero. Similarly, Oscilloscope 2 measures another set of voltage signals as $V_2(t)=\delta V(t)+V_{\text{osc2}}(t)$. Thus, the cross-correlation noise spectrum is

$$S_{\text{cross}}(f) = \frac{2}{t_f} \langle V_1(f) V_2(f)^* \rangle$$
$$= \frac{2}{t_f} \langle |\delta V(f)|^2 + V_{\text{osc1}}(f) V^*_{\text{osc2}}(f) + \delta V(f) V^*_{\text{osc2}}(f) + V_{\text{osc1}}(f) \delta V^*(f) \rangle \quad (6)$$
$$= S_{\delta V}(f),$$

where $V_2(f)^*$ is the complex conjugate of the Fourier transform of $V_2(t)$. In the second step above, the expectations of the last three terms are zero because $\delta V(t)$, $V_{\text{osc1}}(t)$, and $V_{\text{osc2}}(t)$ are independent. A comparison of Eqs. (5) and (6) shows that the cross-correlation noise measurement effectively suppresses the uncorrelated background noise from the experimental instruments, such as the voltage noise of the amplifier and the ADC. Interference signals present in both channels remain correlated and are thus not suppressed by cross-correlation. Examples include low-frequency pickup from AC mains (50/60 Hz and harmonics) or ground noise caused by the voltage drop generated when ground current flows through the ground wire.

## III. INSTRUMENTS

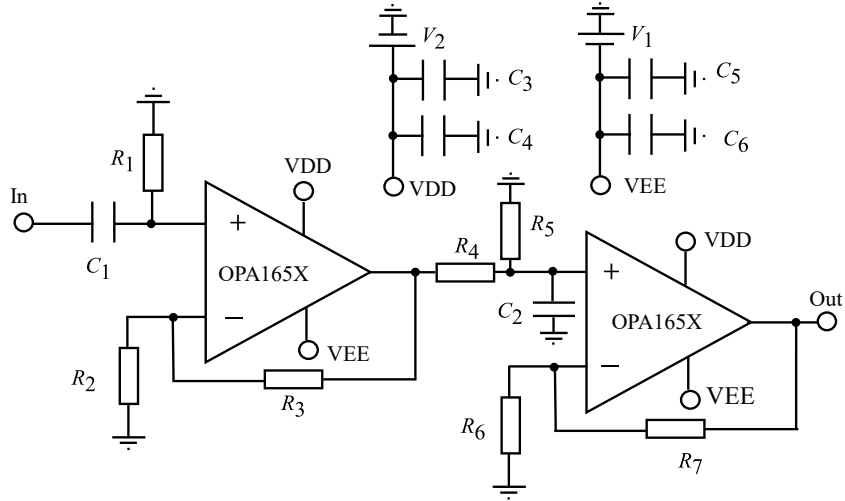

Fig. 2: Circuit schematic of the amplifier, all values of the circuit elements of the setup are given in Table A1.

When acquiring signals with a low-frequency ADC operating within an effective bandwidth below 10 kHz, it is critical that the preamplifier exhibits sufficiently low 1/f noise. For this experiment, the OPA165X [20] was chosen as the preamplifier owing to its 1/f corner frequency below 100 Hz, an input voltage noise of 2.9 nV/$\sqrt{\text{Hz}}$, and an exceptionally low input current noise of 6 fA/$\sqrt{\text{Hz}}$. These characteristics make it particularly suitable for both shot and thermal noise measurements, as further discussed in the following section.

As illustrated in Fig 2, the circuit comprises two cascaded standard non-inverting amplifier stages. A high-value capacitor is connected at the input to block DC current. The first stage provides a voltage gain of $G_1=20\lg(\frac{R_3}{R_2}+1)=26.4$ dB. Between the two stages, an interstage network consisting of resistors $R_4$, $R_5$ and capacitor $C_2$ is inserted. Resistors $R_4$ and $R_5$ form a half voltage divider and, together with $C_2$, they also constitute a low-pass filter with a cutoff frequency of approximately $f=\frac{R_4+R_5}{2\pi R_4 R_5 C_2}\approx 63.6$ kHz. The second stage provides a gain of $G_2=20\lg[(\frac{R_7}{R_6}+1)/2]=34$ dB. Thus, the total gain of the system is 60.4 dB. Since there are two operational amplifiers on each OPA1656 chip, only one chip is needed for this two-stage amplification. This represents a notable advantage for system integration, as shown in the corresponding circuit diagram in Fig. A1. Moreover, the circuit is straightforward, and its standardized gain design allows the gain calibration procedure to be omitted in teaching contexts, making the setup well suited for student-led DIY experiments.

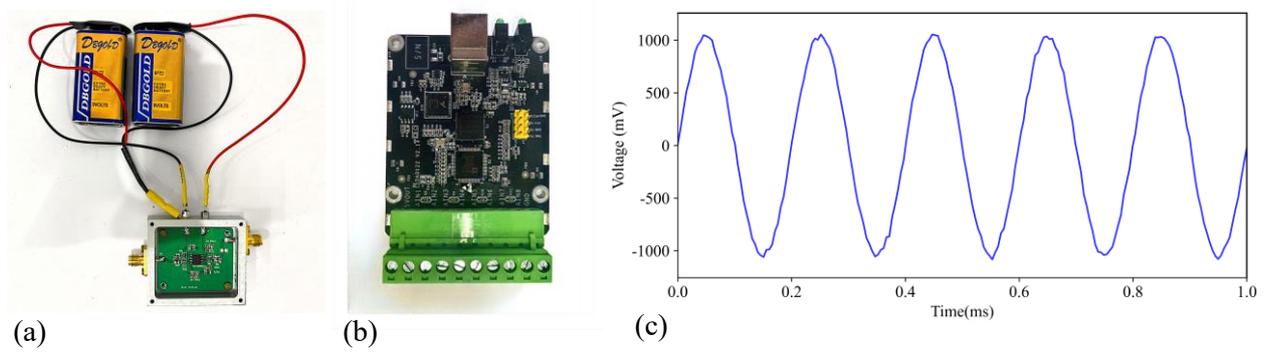

(a) (b) (c)

Fig. 3: (a) Photograph of the amplifier and the power supply used to bias it. (b) Photograph of the ADC module. (c) Time-domain waveform acquired by the ADC module and amplifier.

We utilized conventional 9 V batteries as the experimental power source due to their stable voltage output, low noise level, and additional advantages such as low cost and simple structure. Fig. 3(a) shows the power supply that delivers the required ±9 V bias to the OPA1656 operational amplifier. If rechargeable batteries are used, it is important to check for the presence of any voltage conversion circuit, as this may introduce substantial voltage noise. For long-term operation or educational use, an AC-DC power module can be selected in combination with a voltage regulation module, as illustrated in Fig. A2(a). This approach not only minimizes noise but is also more environmentally friendly.

The experimental setup employed a cost-effective commercial ADC module (AD7606) [21], with up to eight input channels, 16-bit resolution, and a 200 kHz sampling rate. The module is connected to a computer via a USB interface. The manufacturer provides a dedicated C-language driver for data acquisition, which we encapsulated in a Python wrapper to configure the module, set the sampling rate, trigger measurements, and retrieve the voltage time-series data. This software integration allows for flexible, script-controlled operation and straightforward data export for subsequent analysis in Python. To evaluate instrument performance, a 5 kHz, 1 mV sine wave was applied to the amplifier input and sampled by the ADC. As shown in Fig.

3(c), the measured waveform demonstrates a gain of approximately 60 dB, which is consistent with the designed circuit specifications. In fact, within the measurement bandwidth of the experiment, the gain of the signal remain constant. The specific amplifier gain profile is provided in Appendix Fig. A3.

A detailed cost breakdown of the experimental equipment, including the noise circuit boards but excluding the computer used for data processing, is provided in Table 1 below.

| Description | Quantity | Cost (USD) |
| --- | --- | --- |
| Amplifier chip | 2 | ~20.00 |
| Amplifier circuit PCB | 2 | ~3.00 |
| ADC module | 1 | ~30.00 |
| Battery | 6 | ~8.00 |
| Noise circuit PCB | 2 | ~5.00 |
| Aluminum metal shielding box | 4 | ~12.00 |
| Resistor, capacitor, inductor | Several | ~2.00 |
| SMA cable | Several | ~3.00 |
| Total | | ~83.00 |

Table 1: Equipment list and cost estimation.

## IV. EXPERIMENT

### A. Thermal Noise

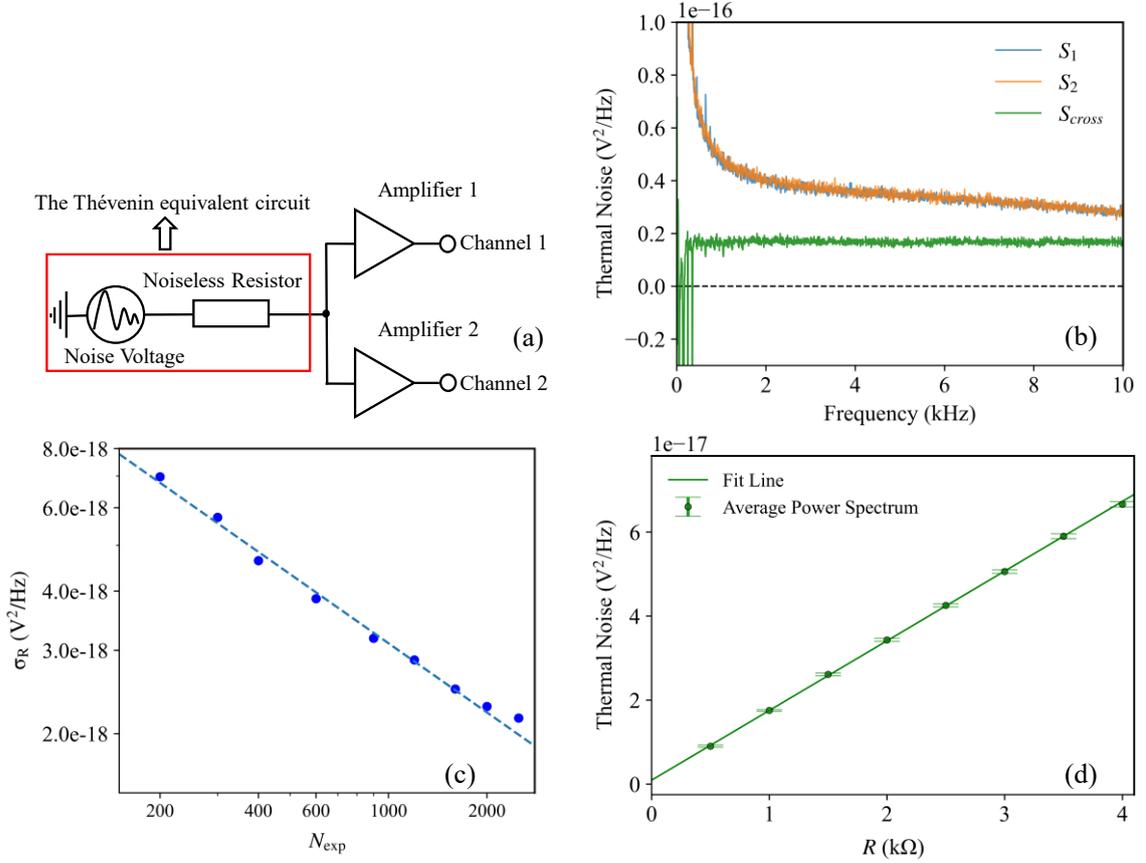

Fig. 4: (a) Schematic diagram of the thermal noise apparatus. (b) Power spectral densities $S_1$ and $S_2$ and, real part $S_{cross}$ of the cross spectral density, were measured at room temperature with a 1 kΩ resistor. (c) Standard deviations of the cross spectral density of a 1kΩ resistor in the frequency range from 500 Hz to 10 kHz are plotted as a function of the number of repetitions; the dotted line follows a power-law dependence and the slope is −0.48. (d) Average cross spectral density plotted as a function of the resistors $R$. The data were best fitted by the equation $S_R=4Rk_BT+S_0$, with the ambient temperature $T$ at 298 K.

At nonzero temperature, a resistor with internal thermal noise can be represented by a Thévenin equivalent circuit [22] consisting of a noiseless resistor in series with a Gaussian noise voltage source, as illustrated in Fig. 4(a). Thermal noise is approximately white, meaning its power spectral density is nearly constant throughout the frequency spectrum. According to the Johnson-Nyquist formula [23–24], the thermal noise from a resistor at kelvin temperature and bandlimited to a frequency band of bandwidth $\Delta f$ has a mean square voltage of $4Rk_BT\Delta f$, where $k_B$ is the Boltzmann constant. If the conductor is a pure resistance, the power spectral density $S_R=4Rk_BT$ is independent of frequency and only depends on the temperature and the resistance value. Thus, by measuring the spectral density of a known resistor at a specified ambient temperature, the Boltzmann constant can be experimentally determined. In our experiment, we measured the cross spectral densities of thin-film chip resistors with resistances ranging determined from 500 Ω to 4 kΩ. The resistors were mounted inside a shielded aluminum enclosure, Fig. A2(b), to minimize external electromagnetic pickup. For each data set, $2^{15}$ (32,768) data points were acquired per single Fourier transform, with the ADC sampling rate fixed at $f_s$=200 kHz, corresponding to a sampling time per iteration of $t_f = N_t / f_s \approx 0.164$ s. The measurement was repeated $N_{exp}$=2000 times, with a total sampling time about 5.5 minutes to improve the signal-to-noise ratio in the frequency domain through averaging. Fig. 4(c) shows the standard deviation $\sigma_R$ of the cross spectral density as a function of $N_{exp}$ on a logarithmic scale. As the number of repetitions increases, the standard deviation exhibits a decreasing trend. According to statistical theory [25], if the variance of the power spectral density for a single measurement is $\sigma^2$, after averaging over $N_{exp}$ repetitions, the variance reduces to $\sigma^2/N_{exp}$, and the corresponding standard deviation decreases as $1/\sqrt{N_{exp}}$. The slope of the dotted line is −0.48, which also corresponds to the square root after taking the logarithm. For $N_{exp}$ =2000 times, A single resistor measurement takes only a few minutes, making it suitable for classroom experiments.

The cross-correlation noise exhibits a constant power spectral density, apart from the interference observed in the low-frequency domain, as shown in Fig. 4(b). The low-frequency 50Hz AC power supply used by laboratory equipment such as computers is very likely the cause of this electromagnetic interference. In the single-channel noise spectrum, obvious 1/f noise is observed below 2 kHz. As previously stated, the 1/f corner frequency of the OPA1655 amplifier is very low, approximately 100 Hz, which suggests that the ADC module is the primary noise source in this frequency range. The amplitude of the single-channel noise spectral density is significantly higher than that of the cross-correlation noise. This discrepancy arises because the experimental setup, including the amplifiers, contributes background noise that is independent between the two channels. At a room temperature of 298 K, the theoretical thermal noise power spectral density for a 1 kΩ resistor is $1.6807×10^{-17}$ V²/Hz. The measured average cross spectral density over the frequency range from 500 Hz to 10 kHz is $1.6567×10^{-17}$ V²/Hz, which is in excellent agreement with theory. These results demonstrate that the cross-correlation technique effectively suppresses uncorrelated noise components. Furthermore, since the single-channel noise does not exhibit white noise characteristics, its average power spectral density varies across different frequency bands. In contrast, the cross-correlation noise power spectral density shows much more stable behavior over the 500 Hz to 10 kHz range.

As shown in Fig. 4(d), a highly linear relationship is obtained between thermal noise and resistance value. The error bars for each point are calculated based on the standard deviation within the same selected frequency band. The fit line goes almost through the origin; the y-intercept $S_0$ is only $1.14×10^{-18}$ V²/Hz, which is an order of magnitude smaller than the thermal noise levels measured, further confirming that the cross-correlation method effectively subtracts instrumental background noise. The value of 298 K was obtained using a common thermometer placed near the experimental apparatus, and the accuracy of this value was not the main focus of the experiment. In fact, there was no power consumption loaded on the resistor during the experiment. Therefore, the temperature deviation was very small. It is conservatively estimated that the impact on the experimental results is within 0.1%. The slope is $(1.6592\pm0.0117)\times10^{-20}$ J, where the error is calculated from the covariance matrix of the weighted least-squares fit. Using this slope and the measured ambient temperature, we obtain finally the Boltzmann constant $k_B = (1.3931\pm0.0098) \times 10^{-23}$ J/K. This result can be compared to the accepted value of $1.3806\times10^{-23}$ J/K, with a final relative uncertainty in our measurement of only about 1.6%.

It is worth noting that the amplifier used in Reference [6] is the ADA4898 [26], which has a very low input voltage noise of 0.9 nV/$\sqrt{Hz}$. This is a recently published study and is primarily designed for shot noise measurement. However, its input current noise is relatively high at 2.4 pA/$\sqrt{Hz}$, substantially exceeding that of the OPA1656 amplifier used in this work, which is only 6 fA/$\sqrt{Hz}$. The measured total output voltage noise is influenced by the characteristics of the amplifier and can be expressed as the sum of three components

$$\delta V_{out}^2(f) = G^2[\delta I_{in}^2(f)R^2 + 4Rk_BT + \delta V_{in}^2(f)], \qquad (7)$$

where $G$ is the amplifier gain, $\delta V_{in}^2(f)$ is its input voltage noise, and $\delta I_{in}^2(f)$ is the amplifier's input current noise, so $\delta I_{in}^2(f)R^2$ is the voltage noise across the resistor $R$ due to the current noise. When the amplifier's input current noise is significant, this quadratic term $\delta I_{in}^2(f)R^2$ can considerably affect the measurement. The input current noise of the amplifier, $\delta I_{in}(t)$, flows through the same resistor $R$ in both measurement channels, generating a common voltage noise that appears identically in two channels. It does not average to zero in the cross-correlation spectrum, unlike the uncorrelated. Consequently, the ADA4898 is not well suited for thermal noise experiments.

## B. Shot Noise

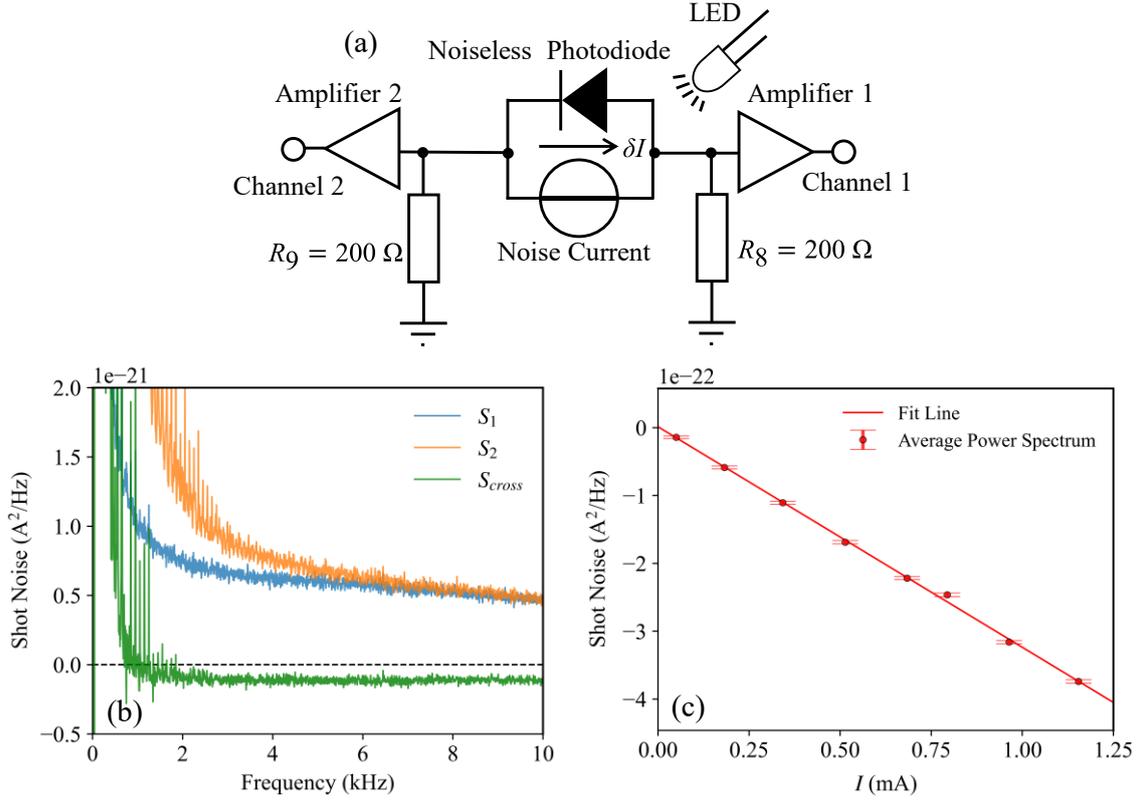

Fig. 5: (a) Schematic diagram of shot noise apparatus. (b) Power spectral densities $S_1$ and $S_2$ and, real part $S_{\text{cross}}$ of the cross spectral density, with the DC photocurrent of 0.78 mA, where the noise circuit and amplifier are both powered by dry batteries. (c) Average cross spectral density of shot noise plotted as a function of the current $I$ and the data were best fitted by the equation $S_I=-2eI+S_0$.

Shot noise in electronic circuits originates from random fluctuations in direct current, which result from the discrete nature of charge carriers. Based on the assumption that the statistics of electron passage follow a Poisson distribution, Schottky derived the spectral noise density as $S_I=2eI$ [27–28]. As illustrated in Fig. 5(a), a photodiode exhibiting shot noise can be represented by a Norton equivalent circuit [29]. This circuit consists of a noiseless photodiode in parallel with a Gaussian noise current source. The flow of noise current $\delta I$ through a resistor $R$ generates a voltage noise with a power spectral density of $S_V=S_I R^2$. This quantity, $S_V$, is what is directly measured in the experiment. In accordance with Kirchhoff's law [30], the voltage noises produced by the noise current across resistors $R_8$ and $R_9$ have opposite polarities Therefore, the cross spectral density between the two channels is expected to be negative. Dividing this value by the square of the known resistance yields the shot noise.

In the experiment, the photocurrent through the diode was adjusted by varying the intensity of the LED, enabling the acquisition of shot noise across different current levels. We adopted the method described in Reference [6] to create the photocurrent; detailed circuit configurations are provided in Appendix Fig. A4. And Fig. 5(b) shows the measurement results of the cross spectral density of shot noise when the photocurrent $I$=0.78 mA. Note that the two single-channel noise exhibit non-ideal behavior at low frequencies. This arises because, in the detailed circuit, capacitor $C_{11}$ effectively acts as a short to ground for high-frequency signals, resulting in the equivalent circuit shown in Fig. 5(a). At low frequencies, however, the impedance of $C_{11}$ becomes significant, breaking the symmetry between the two sides of the circuit. Thus, when calculating the average spectral density, it should start from at least 2 kHz. Following the method applied in the thermal noise experiment, we performed linear regression on the average shot noise against the photocurrent. Fig. 5(c) shows the results, which exhibit a highly linear relationship, with a y-intercept two orders of magnitude smaller than the measured shot noise levels and the standard deviation of each red circle is approximately 1%. The slope of the fit line is $-2e=-(3.2497\pm0.0217) \times 10^{-19}$ C, from which we extracted the electron charge value $e$ as $(1.6249\pm0.0109) \times 10^{-19}$ C. This value can be compared to the accepted value of $1.6022\times10^{-19}$C; we note that the final uncertainty in our result is only about 2%.

## V. CONCLUSIONS

We achieved high-fidelity extraction of the weak thermal and shot noise signals, resulting in measurement accuracies for both the Boltzmann constant ($k_B$) and the elementary charge ($e$) of approximately 1–2%. The system demonstrates several pedagogical advantages. It allows students to directly observe noise spectra and compare single-channel and cross-correlation results. The use of commercially available components and a modular circuit design makes the setup highly accessible, reproducible, and suitable for DIY experimentation in both undergraduate laboratories and advanced high school courses.

## VI. AUTHOR CONTRIBUTIONS

Zitong Peng (**First Author**): Conceptualization, Methodology, Investigation, Software, Validation, Formal analysis, Data Curation, Visualization, Writing-Original Draft.

Jie Zheng (**Second Author**): Conceptualization, Methodology, Supervision, Writing-Review & Editing.

Xiaokai Yue (**Corresponding Author**): Conceptualization, Methodology, Supervision, Funding acquisition, Writing-Review & Editing, Funding acquisitions.

## ACKNOWLEDGMENTS

We would like to express our heartfelt gratitude to all the predecessors for their valuable contributions to the experimental exploration of noise education.

1. Earl, J. A. (1966). Undergraduate experiment on thermal and shot noise. *American Journal of Physics*, *34*(7), 575-579. https://doi.org/10.1119/1.1973117
2. Kittel, P., Hackleman, W. R., & Donnelly, R. J. (1978). Undergraduate experiment on noise thermometry. *American Journal of Physics*, *46*(1), 94-100. https://doi.org/10.1119/1.11171
3. Pruttivarasin, T. (2018). A robust experimental setup for Johnson noise measurement suitable for advanced undergraduate students. *European Journal of Physics*, *39*(6), 065102. https://doi.org/10.1088/1361-6404/aadc8b
4. MIT Advanced Laboratory (8.13) course manual, MIT Physics Department, 2023, https://web.mit.edu/8.13/www/JLExperiments/JLExp43.pdf
5. Spiegel, D. R., & Helmer, R. J. (1995). Shot-noise measurements of the electron charge: An undergraduate experiment. *American Journal of Physics*, *63*(6), 554-559. https://doi.org/10.1119/1.17867
6. Mishonov, T. M., Petkov, E. G., Mihailova, N. Z., Stefanov, A. A., Dimitrova, I. M., Gourev, V. N., ... & Varonov, A. M. (2018). Simple do-it-yourself experimental set-up for electron charge qe measurement. *European Journal of Physics*, *39*(6), 065202. https://doi.org/10.1088/1-3616404/aad3d7.
7. Mishonov, T. M., Serafimov, N. S., Petkov, E. G., & Varonov, A. M. (2022). Set-up for observation thermal voltage noise and determination of absolute temperature and Boltzmann constant. *European Journal of Physics*, *43*(3), 035103. https://doi.org/10.1088/1361-6404/ac5e15.
8. Fink, H. J. (1959). A new absolute noise thermometer at low temperatures. *Canadian Journal of Physics*, *37*(12), 1397-1406. https://doi.org/10.1139/p59-161
9. Pellegrini, B., Basso, G., Fiori, G., Macucci, M., Maione, I. A., & Marconcini, P. (2013). Improvement of the accuracy of noise measurements by the two-amplifier correlation method. *Review of scientific instruments*, *84*(10). https://doi.org/10.1063/1.4823780
10. Kraftmakher, Y. (2002). Correlation analysis with ScienceWorkshop. *American Journal of Physics*, *70*(7), 694-697. https://doi.org/10.1119/1.1475330
11. Pérez, A. T. (2011). Measuring the speed of electromagnetic waves using the cross-correlation function of broadband noise at the ends of a transmission line. *American Journal of Physics*, *79*(10), 1042-1045. https://doi.org/10.1119/1.3607431
12. Kraftmakher, Y. (1995). Two student experiments on electrical fluctuations. *American Journal of Physics*, *63*(10), 932-934. https://doi.org/10.1119/1.18035
13. DiCarlo, L., Zhang, Y., McClure, D. T., Marcus, C. M., Pfeiffer, L. N., & West, K. W. (2006). System for measuring auto-and cross correlation of current noise at low temperatures. *Review of scientific instruments*, *77*(7). https://doi.org/10.1063/1.2221541.
14. Callsen, G., Carmele, A., Hönig, G., Kindel, C., Brunnmeier, J., Wagner, M. R., ... & Arakawa, Y. (2013). Steering photon statistics in single quantum dots: From one-to two-photon emission. *Physical Review B—Condensed Matter and Materials Physics*, *87*(24), 245314. https://doi.org/10.1103/PhysRevB.87.245314
15. Rieke, F., Warland, D., Van Steveninck, R. D. R., & Bialek, W. (1999). *Spikes: exploring the neural code*. MIT press. Spikes
16. Oppenheim, A. V., & Verghese, G. C. (2016). *Signals, systems, and inference*. Pearson. Signals, Systems and Inference, complete notes
17. Youngworth, R. N., Gallagher, B. B., & Stamper, B. L. (2005). An overview of power spectral density (PSD) calculations. *Optical manufacturing and testing VI*, *5869*, 206-216. https://doi.org/10.1117/12.618478
18. J. Loren Passmore, Brandon C. Collings, Peter J. Collings; Autocorrelation of electrical noise: An undergraduate experiment. *Am. J. Phys.* 1 July 1995; 63 (7): 592–595. https://doi.org/10.1119/1.17892
19. Walls, W. F. (1992, May). Cross-correlation phase noise measurements. In *Proceedings of the 1992 IEEE Frequency Control Symposium* (pp. 257-261). IEEE. https://doi.org/10.1109/FREQ.1992.270007.
20. Texas Instruments. (2022). OPA165x high-performance, bipolar-input audio operational amplifiers datasheet (Rev. C) [Data sheet]. OPA165x Ultra-Low-Noise, Low-Distortion,

| | |
|---|---|
| | FET-Input, Burr-Brown™ Audio Operational Amplifiers datasheet (Rev. C). |
| 21 | Analog Devices. (2025). AD7606 8-channel DAS with 16-bit, bipolar input, simultaneous sampling ADC (Rev. G) [Data sheet]. AD7606/AD7606-6/AD7606-4 (Rev.G). |
| 22 | Brittain, J. E. (1990, March). Thevenin's theorem. *IEEE Spectrum, 27*(3), 42. https://doi.org/10.1109/6.48845. |
| 23 | Johnson, J. B. (1928). Thermal agitation of electricity in conductors. *Physical review*, *32*(1), 97. https://doi.org/10.1038/119050c0. |
| 24 | Nyquist, H. (1928). Thermal agitation of electric charge in conductors. Physical review, 32(1), 110. https://doi.org/10.1103/PhysRev.32.110. |
| 25 | Rice, J. A. (2007). *Mathematical Statistics and Data Analysis* (3rd ed.). Thomson/Brooks/Cole. (PDF) Mathematical Statistics and Data Analysis |
| 26 | Analog Devices. (2015). ADA4898-1/ADA4898-2 3 nV/√Hz, low noise, high speed, 18 V, CMOS input rail-to-rail output operational amplifiers (Rev. D) [Data sheet]. ADA4898-1/ADA4898-2 (Rev. D). |
| 27 | Blanter, Y. M., & Büttiker, M. (2000). Shot noise in mesoscopic conductors. *Physics reports*, *336*(1-2), 1-166. https://doi.org/10.1016/s0370-1573(99)00123-4 |
| 28 | Schottky, W. (1918). Über spontane Stromschwankungen in verschiedenen Elektrizitätsleitern. *Annalen der physik*, *362*(23), https://doi.org/ 10.1002/andp.19183622-304 |
| 29 | Norton, E. L. (1926). Design of finite networks for uniform frequency characteristic. Bell Labs, Murray Hill, NJ, USA, Tech. Rep. TM26-0-1860. Nortonmemo.pdf |
| 30 | Alexander, C., & Sadiku, M. (2006). *Fundamentals of electric circuits* (3rd rev. ed., pp. 37–43). McGraw-Hill. Fundamentals of Electric Circuits, 7th edition - MATLAB & Simulink Books |
| 31 | Analog Devices. (2015). LT3045, ultralow noise, ultrahigh PSRR RF linear regulator (Rev. D) [Data sheet]. LT3045 (Rev. D). |
| 32 | Analog Devices. (2005). LT3015, ultralow noise, ultrahigh PSRR RF linear regulator (Rev. E) [Data sheet]. LT3015-1.5A, Low Noise, Negative Linear Regulator with Precision Current Limit. |
| 33 | Vishay Intertechnology, Inc. (2007). BPW34, BPW34S, BPW34F, BPW34FS Silicon PIN Photodiode (Rev. D) [Data sheet]. bpw34.pdf. |